\documentstyle[twoside,fleqn,epsfig,espcrc2]{article}
\def\beq{\begin{equation}}
\def\eeq{\end{equation}}
\def\bea{\begin{array}}
\def\eea{\end{array}}
\def\beqa{\begin{eqnarray}}
\def\eeqa{\end{eqnarray}}

\def\u1{{U(1)}}
\def\su2{{SU(2)}}
\newcommand{\re}{\relax{\rm I\kern-.18em R}}

\def\pl{{{\cal P}_\infty}}

\newcommand{\AmS}{{\protect\the\textfont2
  A\kern-.1667em\lower.5ex\hbox{M}\kern-.125emS}}

\def\frac#1#2{ {{#1} \over {#2} }}

\def\beq{\begin{equation}}
\def\eeq{\end{equation}}
\def\re#1{(\ref{#1})}
\def\L{\Lambda}

\def\b{\beta}

\def\g{\gamma}
\def\as{\alpha_{\sf s}}

\def\np#1#2#3{Nucl.\ Phys.\ B#1 (19#3) #2}
\def\pl#1#2#3{Phys.\ Lett.\ #1B (19#3) #2}
\def\pr#1#2#3{Phys.\ Rev.\ D #1 (19#3) #2}

\newcommand{\ban}        {\begin{eqnarray*}}
\newcommand{\ean}        {\end  {eqnarray*}}

\title{$\beta$-function, Renormalons and the Mass Term from 
Perturbative Wilson Loops
\vskip-2.4cm\hfill\small UPRF-99-15; BICOCCA-FT-99-29\vskip2.4cm
}
\author{G.\ Burgio\address{Dipartimento di Fisica, Universit\`a di Parma
        and INFN, Gruppo Collegato di Parma, Italy},
        F.\ Di Renzo$^{\rm a}${}$\thanks{Presented by the second author at 
	Lattice '99}$,
        M.\ Pepe\address{Dipartimento di Fisica, Universit\`a di Milano
        and INFN, Sezione di Milano, Italy}
        and L.\ Scorzato$^{\rm a}$}

\begin{document}
\begin{abstract}
Several Wilson loops on several lattice sizes are computed in Perturbation 
Theory via a stochastic method. Applications include: Renormalons, 
the Mass Term in HQET and (possibly) the $\beta$-function.
\end{abstract}

\maketitle

\section*{Outlook}
Wilson Loops (WL) were the historic playground (and success \ldots) of 
Numerical Stochastic Perturbation Theory (NSPT) for Lattice Gauge Theory 
\cite{NSPT}. 
Having by now an increased computing power available, we are 
computing high perturbative orders of various WL 
on various lattice sizes in Lattice $SU(3)$. Physical motivations range over 
a variety of issues (not every one within reach, at the moment): 
Renormalons and Lattice Perturbation 
Theory (LPT), the Mass Term in Heavy Quark Effective Theory (HQET) and the 
Lattice $\beta$-function. 

\section{Renormalons and LPT}
In ~\cite{8L} WL of sizes $1\times1$ and $2\times2$ were computed in 
LPT via NSPT up to $\beta^{-8}$ order. The expected Renormalon 
contribution was found according to the formula ($Q=a^{-1}$)
\beq\label{REN}
W_0^{\rm ren} = {\cal N} \;  
\int^{Q^2}_{r\Lambda^2}\;
\frac{k^2\,dk^2}{Q^4}
\;\as(k^2)
\,.
\eeq
which is fixed by dimensional and Renormalization Group considerations. 
\re{REN} can be cast in a form from which a power expansion can be easily 
extracted (two loop asymptotic scaling is assumed)  
\ban
z=z_0(1-\frac{\as(Q^2)}{\as(k^2)}) &,& z_0 = \frac{1}{3b_0} \\
z_{0-}=z_0(1-\frac{\as(Q^2)}{\as(r\Lambda^2)}) &,& \gamma = 2 \frac{b_1}{b_0^2} 
\ean
\ban
W_0^{\rm ren}  &=& C' \int_{0}^{z_{0-}} dz\;e^{-\b z}
\;(z_0-z)^{-1-\g} \\ 
&=& \sum_{\ell=1}\b^{-\ell}\{C'\Gamma(\ell+\g)z_0^{-\ell}
+{\cal O}(\L^4/Q^4)\}
\,,
\ean
Actually \re{REN} refers to some continuum scheme. On a finite lattice 
(which is what NSPT needs) one has to deal with
\beq\label{FLREN}
W_0^{\rm ren}(s,N) = C \;  
\int^{Q^2}_{Q_0^2(N)}\;
\frac{k^2\,dk^2}{Q^4}
\;\as(s k^2)
\,.
\eeq
The factor $s$ in the argument of $\as$ is in charge of the 
lattice-continuum matching and is expected to be of the order $s \sim 
\left( \frac{\Lambda_{lat}}{\Lambda_{cont}} \right)^2$ with respect to some 
continuum scheme. An explicit IR cut-off is present, dependent on the 
lattice size ($Q_0(N) = 2\pi(Na)^{-1}$, where 
$N$ is the number of points in any direction). \re{FLREN} 
results in a new power expansion
\beq\label{COEFF}
W_0^{\rm ren}(s,N) = 
\sum_\ell \, C_\ell^{\rm ren}(C,s,N) \, \beta^{-\ell}
\eeq
whose coefficients $C_\ell^{\rm ren}(C,s,N)$ are given in terms of incomplete 
$\Gamma$-functions. In \cite{8L} (via a slightly different, equivalent 
formalism) it was shown that this renormalon contribution can account for 
the growth of the first $8$ coefficients in the pertubative expansion of the 
plaquette
\ban
W^{1 \times 1} = \sum_{\ell=1}^8 c^{1 \times 1}_\ell \, \beta^{-\ell}
\ean
{\em i.e.} convenient values $C^*$ and $s^*$ were fitted so that 
$c^{1 \times 1}_\ell$ were recognized to be asymptotically the same as 
$C_\ell^{\rm ren}(C^*,s^*,8)$ (in \cite{8L} computations were performed on 
a rather small $N=8$ lattice);  $C_\ell^{\rm ren}(C^{**},s^*,8)$ 
(with a different choice for $C$) were also shown to fit the expansion of 
$W^{2 \times 2}$.

Control on this (renormalon) perturbative contribution is crucial in the 
analysis presented in \cite{L2}: once its resummation is subtracted from 
Monte Carlo data for the plaquette, one is left with a ${\cal O}(\L^2/Q^2)$ 
contribution rather than the expected ${\cal O}(\L^4/Q^4)$. In order to 
trust this remarkable result, one would like to further test the asymptotic 
formula \re{COEFF} by going to even higher orders in the perturbative 
expansion on various lattice sizes. 

By now we know the expansion of the basic plaquette up to order 
$\beta^{-10}$ on both $N=8$ and $N=24$ lattices. We do not present here 
the definitive results (which will be published soon elsewhere), but 
we do show how well the new results are described by 
$C_\ell^{\rm ren}(C^*,s^*,N)$ ({\em i.e.} by the values for $C$ and $s$ 
that were fitted in \cite{8L,L2}). In the figure 
the expected $C_\ell^{\rm ren}(C^*,s^*,24)$ are plotted together with the 
computed $c^{1 \times 1}_\ell$ ($\ell=1 \ldots 10$) for the $N=24$ lattice. 
The renormalon contribution is indeed there just like described in 
\cite{8L}. Work is in progress to perform the whole analysis of \cite{L2} 
on the $N=24$ lattice. 

\begin{figure}[htb]
\centerline{\epsfig{file=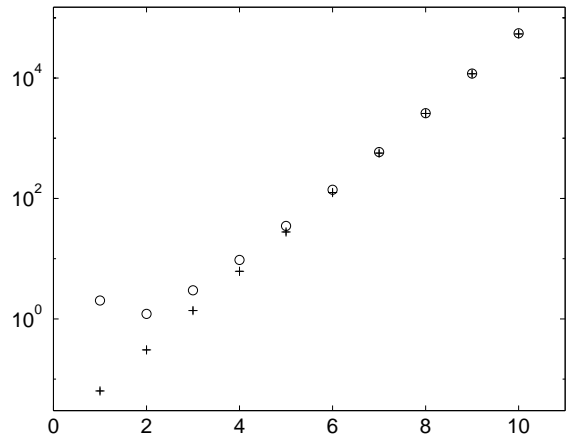,width=7.5cm}}
\caption{$c^{1 \times 1}_\ell$ obtained on a $N=24$ lattice (circles) 
and $C_\ell^{\rm ren}(C^*,s^*,24)$ (crosses) vs perturbative order ${\ell}$.}
\end{figure}

\section{The Mass Term in HQET}
Consider WL of various sizes (in particular square $L \times L$ loops). 
From their renormalization properties \cite{Dots} one has to expect
\beq
W(L) = \exp^{-4M\,L}\;w(L)
\eeq
The first factor is the exponential of the perimeter times the Mass Term 
one has to deal with in HQET (an additive, linearly divergent mass 
renormalization). $w(L)$ contains logarithmic divergences, in particular 
only those connected to the coupling renormalization is there is no 
``corner'' (which is not the case on the lattice). 
As stressed in Hashimoto's review 
at this conference \cite{Hash} (after \cite{GuidoChris}), the mass term 
is a fundamental building block in a renormalon-safe determination of 
the b-quark mass from the lattice. 

The mass term could of course be determined from the heavy quark 
propagator. The determination we are going to report on (again after 
\cite{GuidoChris}) goes through the computation of various WL and is 
well suited for NSPT as it is gauge invariant \cite{FERMILAB}. 
By computing perturbative expansions of WL of various sizes 
\ban
W(L) = 1 - W_1(L)\, \beta^{-1} - W_2(L)\, \beta^{-2} + \ldots
\ean
and therefore 
\ban
- \log W(L) &=& W_1(L)\, \beta^{-1} + \\ 
& & \left( W_2(L)+\frac{1}{2}W_1(L)^2 \right)\beta^{-2} + \ldots
\ean
one can extract from each order the leading (linear) behaviour in $L$, 
that is mass term
\beq
4\,M\,L = 4 \left( M_1\, \beta^{-1} + M_2\, \beta^{-2} + \ldots \right) L
\,.
\eeq
At the moment we have got results for square loops on a $(N=24)^4$ 
lattice. Results on bigger lattices ($N=28,32$) and rectangular loops 
are expected to come soon. They will be crucial to control finite size 
effects and subleading (logarithmic) contributions. As $M_1$ and $M_2$ 
are already known, they have to be recovered. Consider for example $M_1$, 
which has to be recovered by fitting $W_1(L)$ to 
\ban
W_1(L) = 4\,M_1\,L + k_1 \log L + k_0 
\ean
The analytical result is $M_1 = 1.01$ and we get $M_1 = 1.00 \pm 0.02$. 
From \cite{GuidoChris} one learns the second coefficient $M_2 = 2.54$, 
while from our fits we get $M_2 = 2.43 \pm 0.15$. The errors we quote 
depend on both finite size effects and fitting subleading contributions. 
Note that since $W_3(L)$ contains a $(\log L)^2$ contribution, the latter 
point is even more crucial in the determination of $M_3$. At the moment we 
can pin down a preliminary number $M_3 = 8 \pm 1$ which will turn in a 
definite number as soon as we get definite results not only for square 
loops, but also for ratios of rectangular loops. 

\section{The Lattice $\beta$-function}
We now turn to describe what could possibly be another application of 
our perturbative computations. Ratios of WL combined in such a way that 
the corner and mass contributions cancel out were introduced several 
years ago by Creutz \cite{Creutz} in order to study (among other things) 
the non-perturbative Lattice $\beta$-function. 

Consider rectangular WL of sizes $(L,L)$, $(2L,L)$, $(2L,2L)$ on a 
$(4L)^4$ lattice and form the ratio
\ban
{\cal R}(L) &=& \frac{W(L,L)\,W(2L,2L)}{W(L,2L)^2} \\
&=& 1 + c_1(L)\, \alpha_0 + c_2(L)\, \alpha_0^2 + \ldots
\ean
where $\alpha_0$ is the bare lattice coupling. 
The only scale is $L$ and the only renormalization needed is 
that of the coupling, so that (for example)
\beq
\alpha_{{\cal R}}(L) = \frac{1}{c_1(L)} \log({\cal R}(L))
\eeq
is supposed to be an acceptable coupling running with L. Since the 
pertubative matching between coupling in different schemes 
\ban
\alpha_1(s\mu) = \alpha_2(\mu) + d_1(s)\, \alpha_2(\mu)^2 + \ldots 
\ean
contains all the informations about the perturbative $\beta$-functions 
in both schemes, one could in principle study the perturbative Lattice 
$\beta$-function by 
computing the perturbative expansion of $\alpha_{CR}(L)$ in $\alpha_0$ 
at different values of $N$, where $L=Na$ ({\em i.e.} by computing 
different WL on different lattice sizes). 

Within this application the big issue is accuracy. Due to large 
cancellations of the mass contribution, a fraction of {\em per mille} 
accuracy on the $c_i(L)$ (which is within reach) straight away 
degenerates when one computes for example $d_1(s)$ (still we were able to 
recover $b_0$, the first universal coefficient of the $\beta$-function). 
In view of this, it is quite unlikely that one can attain the terrific 
accuracy which should be needed to get the first unknown information. 
One should most probably look for a smarter definition of the coupling.

\section*{Conclusions}
Preliminary results were reported, coming from the computations of 
perturbative expansions of various WL. The basic plaquette is now known 
up to $\beta^{-10}$ order and the results on Renormalons \cite{8L} 
are indeed to be trusted. Work is in progress to gain further confidence 
in the results of \cite{L2} as well. A preliminary results on third order 
in the computation of the mass term has been reported and a definite 
result will be published soon. In principle also the perturbative lattice 
$\beta$-function could be studied via WL in NSPT, even if the accuracy 
needed to get the first unknown result is quite unlikely to be attained.

\end{document}